\documentclass[conference]{IEEEtran}
\IEEEoverridecommandlockouts
\usepackage{cite}
\usepackage{amsmath,amssymb,amsfonts}
\usepackage{graphicx}
\usepackage{textcomp}
\usepackage{xcolor}

\usepackage{algorithm}
\usepackage{algpseudocode}
\usepackage{algpascal}

\def\BibTeX{{\rm B\kern-.05em{\sc i\kern-.025em b}\kern-.08em
    T\kern-.1667em\lower.7ex\hbox{E}\kern-.125emX}}
\begin{document}

\title{\huge RSS-Based UAV-BS 3-D Mobility Management via Policy Gradient Deep Reinforcement
Learning
\thanks{This work was supported by Huawei Canada Co., Ltd.}
}

\author{\IEEEauthorblockN{ Mohammad G. Khoshkholgh}
\IEEEauthorblockA{\textit{Carleton University}
\\ghadir@sce.carleton.com} \and \IEEEauthorblockN{Halim
Yanikomeroglu} \IEEEauthorblockA{\textit{Carleton University}\\
halim@sce.carleton.ca} }

\maketitle

\begin{abstract}
We address the mobility management of an autonomous UAV-mounted
base station (UAV-BS) that provides communication services to a
cluster of users on the ground while the geographical
characteristics (e.g., location and boundary) of the cluster, the
geographical locations of the users, and the characteristics of
the radio environment are unknown. UAV-BS solely exploits the
received signal strengths (RSS) from the users and accordingly
chooses its (continuous) 3-D speed to constructively navigate,
i.e., improving the transmitted data rate. To compensate for the
lack of a model, we adopt policy gradient deep reinforcement
learning. As our approach does not rely on any particular
information about the users as well as the radio environment, it
is flexible and respects the privacy concerns. Our experiments
indicate that despite the minimum available information the UAV-BS
is able to distinguish between high-rise (often non-line-of-sight
dominant) and sub-urban (mainly line-of-sight dominant)
environments such that in the former (resp. latter) it tends to
reduce (resp. increase) its height and stays close (resp. far)
to the cluster. We further observe that the choice of the reward
function affects the speed and the ability of the agent to adhere
to the problem constraints without affecting the delivered data
rate.
\end{abstract}


\section{Introduction}
Using unmanned aerial vehicles (UAVs), also known as drones,
benefits many applications including package delivery, search and
rescue, infrastructure monitoring, law enforcement, and the like
\cite{AlzenadPlacement}, \cite{SkyAccessing}, \cite{UAVchallengesZhang}. Due to growing
popularity and low cost, UAVs are getting an increased attention
in the telecommunications sector to address on-demand data
delivery, flexible backhauling, data harvesting, IoT applications,
and caching \cite{SkyAccessing}, \cite{UAVchannel}. For example,
due to their maneuverability, UAVs are exploited to enhance the
performance of wireless communications via optimally deploying
them as aerial (flying) base stations (UAV-BSs). This technique is
shown to be particularly effective for those scenarios involving
high (temporarily-lived) localized traffic surges, e.g. caused by
crowded events, as well as network failure.

Our focus in this paper is to maximize the transmission capacity
of UAV-BS to serve a cluster of users on the ground. This is a
challenging problem given that the 3-D location of UAV-BS and
geographical location of users on the ground affect signal
propagation, and thus the transmission data rate, in a compound
manner. Conventionally, to solve this optimization problem it is
assumed that 1) users are located in a cluster that has particular
mathematical features (for example a circular disk), which its
boundaries are known, 2) accurate knowledge of the channel is
available, and 3) the geographical locations of users are known,
see, e.g., \cite{Al-Hourani}, \cite{BorPlacement}, and
\cite{AlzenadPlacement}. In fact, often the optimal placement of
UAV-BS assumes that the channel can be predicted using a
(relatively) straightforward model that, in the core, exploits the
users' locations along with the knowledge of the radio environment
(sub-urban \emph{vs.} urban areas). Such a knowledge is used to
fine-tune the path-loss and the shadowing parameters, and
consequently to convert the original problem into an equivalent
optimization problem which its objective function as well as
constraints are functions of the location of UAV-BS. In reality,
the information regarding the users' locations may not be
available due to privacy issues or simply the lack of such
knowledge.
Given that the solutions that are not intrusive regarding to the
private information of the users is practically valuable, we,
therefor, promote solutions based on deep reinforcement learning (DRL)
\cite{Sutton} to learn the navigation of UAV-BS in order to
optimize the capacity.

In \cite{LandingSpot}, deep Q-learning network (DQN) is used to
design the trajectory of an autonomous UAV-BS without any explicit
information about the environment. In order to increase the
service time the use of landing spots is also promoted. Assuming a
fixed altitude of the UAV-BS during the navigation, the action
space of the UAV-BS is reduced to 8 movement directions. To
minimize the mission completion time subject to maintaining good
connectivity with the cellular network a temporal difference based
DRL solution is suggested in \cite{UAVconnectedPath}. The
algorithm relies only on the raw signal strength as input. It is
assumed that the UAV flies with fixed speed, therefore the agent
needs only to adjust the direction of the UAV. Note that in all of
these works its also assumed that the destinations are known to
UAV prior to the start of the mission, which may not be the case
for many scenarios.

To address the lack of geographical information of users, work of
\cite{rss} discusses the use of UAV for search and rescue
applications. Authors use UAV for locating a user merely by
receiving the received signal strength (RSS) in an indoor
environment using DQN. It is shown that DQN solution is
competitive with the location-based solution, emphasizing the
power of model-free DRL. In \cite{UAVaccessNoLocation}, DQN is
used to provide connectivity via proper UAV placement in an urban
environment when the location of the end user is unknown. The
algorithm uses the signal-to-interference-and-noise ratio
measurements and exploits 3D map of the topology in order to
account for the scatterers and blockages. In both papers, the
action dimension of the UAV is limited to 4---up, down, left,
right---and the speed is kept fixed. In reality, the mobility of
the UAV-BS is in a continuous 3-D space, which requires a more
sophisticated solutions. Also extracting information from a 3-D
map via reconstruction of the environment can be costly (although
very effective). We hence use only RSS values to navigate the
UAV-BS.

In particular, we adopt trust region policy optimization (TRPO)
algorithm which is a policy gradient DRL to learn the navigation
in continuous 3-D space. Our experiments indicate that UAV-BS
differentiates high-rise (often non-line-of-sight dominant) from
sub-urban (mainly line-of-sight dominant) environments. In effect, while
in the former it tends to reduce
its height and stays closer to the cluster, in the latter it attempts to increase its height and keeps distance to the cluster. We also
demonstrate that the choice of the reward function can affect the
speed and the ability of the agent to adhere to the problem
constraints without affecting the delivered data rate. Last, as our approach does not rely on any particular information
about the users as well as the radio environment, it is flexible
and respects the privacy concerns.

\section{Problem Formulation}
Our main focus is on 3-D navigation of the UAV-BS for providing
communication services to a number of users that are geographically
clustered (a.k.a. the area of interest). The UAV-BS receives RSS
information from the users and accordingly adjusts its location
via modifying its speed $\boldsymbol{v}\in \mathbb{R}^3$ where
$\|\boldsymbol{v}\|=v\in [v_{\min}, v_{\max}]$ m/s. The final goal is to
improve the transmitted data rate. The UAV-BS should stay in the
search area, which is assumed to be a large area with radius
$D_{search}$, during the service time. The maximum and minimum
allowable height values (in meters) that the agent must respect is
$H_{\max}$ and $H_{\min}$, respectively, which are imposed by the
regulator. We consider a time-slotted model in which at the start
of each time slot $t$ the agent chooses a new speed
$\boldsymbol{v}$ and keeps moving by that speed in the chosen
direction unless otherwise it violate the boundaries. The speed is
selected based on the received RSS information. We assume that the
agent equally divides the time slots into $K$ (the number of users) equal parts and
schedules each user in each of them with the transmission power
$P/K$, where $P$ is its instantaneous transmission power budget
(per time slot). We also assume that the uplink channel (between
users and UAV-BS that is dedicated to RSS) and downlink channel (between
UAV-BS and users for data transmission) are frequently
multiplexed. However, further information can be extracted from
RSS information for provisioning a better scheduling and power
allocation schemes, which is left for the future investigation.
The UAV-BS' antenna is directional with beam-width $w$, the
main-lobe antenna gain $G$, and side-lob antenna gain $g$ where
$G\gg g$.

Because the signal strength is a function
of environmental factors such as distance between the agent and
the users, radio environment type (sub-urban versus high-rise),
and the antenna beam-width of the UAV-BS, the agent needs to learn
how to navigate in order to improve the quality of received
signals as well as the transmission data rate. In general, it is
too complex to accurately model such a relationship due to complex
nature of the environment and mobility of the UAV-BS. As a remedy, we
adopt model-free DRL solutions to tackle the involved complexity
of the problem and to effectively deal with the lack of model.

\section{Policy Gradient DRL}
The action of the UAV-BS is its speed in 3-D space, which belongs
to continuous control. Here, we firstly provide a brief
introduction to DRL. We then elaborate on TRPO to handle the
navigation of the UAV-BS.

\subsection{A Brief Introduction to Continuous DRL}
In continuous DRL the agent (UAV-BS), operating in an uncertain
environment with the continuous state and action spaces, interacts
with the environment in a sequential style to learn an optimal
policy (3-D speed) \cite{Lillicrap}. In each interaction the agent
takes an action $\boldsymbol{a}_t\in \mathbb{R}^B$ ($B$ is the
action dimension) based on its observation of the environment
state $\boldsymbol{s}_t\in\mathbb{R}^S$ ($S$ is the dimension of
the state space), which leads the agent to the new state
$\boldsymbol{s}_{t+1}$ upon on collecting the bounded reward
$r_t\in \mathbb{R}$. The policy guides the agent to what action
should be taken in a certain state in order to maximize the reward
via maximizing the aggregate (discounted) expected reward
\cite{Sutton}
\begin{equation}\label{ExpectedReturn}
J(\pi) = \mathbb{E}_{\pi}\sum_{t}\gamma^tr_t(\boldsymbol{s}_t,
\boldsymbol{a}_t)
\end{equation} by finding an optimal policy
$\pi_{\boldsymbol{\theta}}(\boldsymbol{a}_t|\boldsymbol{s}_t)$ (or
for short $\pi_{\boldsymbol{\theta}}$) where $\boldsymbol{\theta}$
are the parameters of the associated DNN\footnote{For given policy
$\pi$, the \emph{state-value function} $V^{\pi}(\boldsymbol{s}_t)$
measures the expected discounted reward from state
$\boldsymbol{s}_t$ via $V^{\pi}(\boldsymbol{s}_t) =
\mathbb{E}_{\boldsymbol{a}_t,
\boldsymbol{s}_{t+1},...}\sum_{t'\geq
t}\gamma^{t'-t}r_{t'}(\boldsymbol{s}_{t'}, \boldsymbol{a}_{t'})$.
The \emph{Q-function} is similarly defined as
$Q^{\pi}(\boldsymbol{s}_t, \boldsymbol{a}_t) =
\mathbb{E}_{\boldsymbol{s}_{t+1},\boldsymbol{a}_{t+1}...}
\sum_{t'\geq t}\gamma^{t'-t}r_{t'}(\boldsymbol{s}_{t'},
\boldsymbol{a}_{t'})$, which is the state-value function for a
given action.}. Parameter $\gamma\in(0, 1]$ is the discount factor
prioritizing short-term rewards and the expectation is on the
policy $\pi$ as well as the stochastic environment dynamics. In
this paper, we focus on stochastic policies by which the DNN
deterministically maps the state to a vector that specifies a
distribution over the action space (i.e.,
$\boldsymbol{a}_t\sim\pi_{\boldsymbol{\theta}}$). To learn the
policy we adopt policy gradient methods in which the gradient
descent with respect to the average return (\ref{ExpectedReturn})
is adopted \cite{Sutton}
\begin{equation}\label{g}
\nabla_{\boldsymbol{\theta}}J(\boldsymbol{\theta})= \boldsymbol{g}
= \mathbb{E}_{\pi_{\boldsymbol{\theta}}}
\sum\limits_{t}\nabla_{\boldsymbol{\theta}}\log\pi_{\boldsymbol{\theta}}(\boldsymbol{a}_t|\boldsymbol{s}_t)A_{\boldsymbol{\theta}}(\boldsymbol{s}_t,
\boldsymbol{a}_t).
\end{equation} Here we use the case that the policy gradient
is formulated through the \emph{advantage function}
$A_{\boldsymbol{\theta}}(\boldsymbol{s}_t, \boldsymbol{a}_t)$,
which is the subtraction of the Q-function and state-value
function: $A_{\pi}(\boldsymbol{s}_t, \boldsymbol{a}_t) =
Q_{\pi}(\boldsymbol{s}_t, \boldsymbol{a}_t) -
V_{\pi}(\boldsymbol{s}_t)$.
In practice, (\ref{g}) should be estimated over a batch of data
collected from the current policy via Monte Carlo technique
(sample based estimate of the policy gradient)\footnote{ In the
rest of this paper, we use symbol $\hat{x}$ as the MC estimation
of quantity $x$.}. The agent iteratively collects data
$(\boldsymbol{s}_t, \boldsymbol{a}_t, r_t, \boldsymbol{s}_{t+1})$,
estimates the gradient of the policy, updates the policy, and then
discards the data. This is basically the policy gradient of
vanilla policy gradient (VPG). In practice, VPG algorithm is not
sample efficient as it needs the agent to takes many samples from
the environment, is brittle in convergence, and suffers from high
variance. A very effective way to deal with these issues is via
imposing a constraint on the policy update, which is the core idea
of TRPO.

\vspace{-.5cm}
\subsection{Trust Region Policy
Optimization (TRPO)}
\subsubsection{Background} To stabilize VPG
algorithm, besides learning the policy it is recommended to learn
a value function \cite{GeneralizedAdvantage}---also known as
\emph{actor-critic} technique. In actor-critic approach a DNN,
called the \emph{actor} or the policy net $\pi_{\boldsymbol{\theta}}$, updates the policy
while another DNN, called the \emph{critic} or the value net $V_{\boldsymbol{\omega}}(\boldsymbol{s}_t)$,
updates the value's parameters denoted by
$\boldsymbol{\omega}$.
The state is feed to both policy network and value network. From
the value network the advantage value $A_{\boldsymbol{\theta}}(\boldsymbol{s}_t,
\boldsymbol{a}_t)$ is estimated. The policy
network provides a distribution over the action in continuous
dimension. It is customary to choose an expressive distribution
such as Gaussian distribution. The output of the policy network
calculates the mean value of this distribution. Note that we do
not need to calculate the standard deviation of the distribution,
as it is calculated form the heads of the policy network. This
approach is shown to stabilize the learning
procedure of the policy network.

Regarding the update of the policy net $\pi_{\boldsymbol{\theta}}$, it is beneficial to ensure that the
gradient ascent does not fail to take the steepest ascent
direction in the metric of parameter space without \emph{too much}
divergence from the current policy. The TRPO algorithm fulfills
this goal by imposing Kullback-Leibler (KL)
divergence\footnote{For probability distributions $P$ and $Q$ over
a given random variable the KL divergence is defined as
$D_{KL}(P||Q)=\mathbb{E}_{P}[\log\frac{P}{Q}]$.} constraint on the
size of policy update at each iteration \cite{TRPO}. Recalling that the policy is stochastic, KL
divergence is a natural choice as it quantifies the closeness of
two probability distributions. In TRPO a \emph{surrogate objective
function} is considered as an estimate of the average return
$J(\pi_{\boldsymbol{\theta}})$, so that in each iteration the
following optimization problem needs to be solved:
\begin{eqnarray}
\mathcal{O}:
&&\hspace{-.5cm}\mathrm{Maximize}_{\boldsymbol{\theta}}\hspace{.3cm}
\mathbb{E}_{\pi_{\boldsymbol{\theta}_k}}
  \left[\frac{\pi_{\boldsymbol{\theta}}(\boldsymbol{a}|\boldsymbol{s})}{\pi_{\boldsymbol{\theta}_k}(\boldsymbol{a}|\boldsymbol{s})}{A}_{\boldsymbol{\theta}_k}(\boldsymbol{s}, \boldsymbol{a})\right]\label{Objective} \\
 \mathrm{s.t.}\hspace{.2cm}&& \hspace{-.6cm} \mathbb{E}_{s\sim \pi_{\boldsymbol{\theta}_k}}\left[D_{KL}
 (\pi_{\boldsymbol{\theta}_k}(.|\boldsymbol{s})||\pi_{\boldsymbol{\theta}}(.|\boldsymbol{s}))\right]\leq
 \delta_{KL}.\label{KL_constraint}
\end{eqnarray}

In short, what this
optimization problem is targeting is to update the current policy
$\pi_{\boldsymbol{\theta}_k}$ via finding the new policy
$\pi_{\boldsymbol{\theta}}$ by maximizing an scaled advantage
function. The constraint, which is called \emph{trust region
constraint}, is KL divergence constraint between the current
policy and the new policy. Thus, under TRPO algorithm the
candidate policy should not be far from the current policy while
it improves the surrogate objective function.

In this form the optimization problem $\mathcal{O}$ is not
computationally affordable. An approximate version of the original
optimization problem is then used:
\begin{eqnarray}
\tilde{\mathcal{O}}:
&&\hspace{-.5cm}\mathrm{Maximize}_{\boldsymbol{\theta}} \hspace{.3cm} {\boldsymbol{g}}^T(\boldsymbol{\theta}-\boldsymbol{\theta}_k) \\
 \mathrm{s.t.}\hspace{.4cm}&& \hspace{-.4cm} (\boldsymbol{\theta} -
 \boldsymbol{\theta}_k)^T{F}_{\boldsymbol{\theta}_k}(\boldsymbol{\theta} - \boldsymbol{\theta}_k)\leq
 \delta_{KL}.
\end{eqnarray}
where the objective function is the first-order approximation of
the surrogate objective function and the constraint is the
second-order approximation of the KL divergence constraint
(\ref{KL_constraint}). Here ${\boldsymbol{g}}$ is the policy
gradient and ${F}_{\boldsymbol{\theta}_k}$ is the Fisher
information matrix (FIM) associated to the average KL divergence
at the current policy $\boldsymbol{\theta}_k$ \cite{TRPO}.

\subsubsection{Algorithm} Algorithm \ref{TRPO} provides the steps of TRPO algorithm.
TRPO has an outer loop indexed by $l=1, 2,\ldots, L$. For each
iteration $l$, the policy is fixed allows the agent to take
actions and collect new bach of data. The iteration comprises of
an inner loop indexed by $n$ with length $N$ (the number of
transitions which also known as batch size), each of which
associated with an episode with length $T$. Using the collected
transitions the advantage function, gradient, and FIM are
estimated via Monte Carlo technique, which are used to update the
policy network and value network.

\alglanguage{pseudocode}
\begin{algorithm}
\caption{TRPO}\label{TRPO}
\begin{algorithmic}[1]
 \State \footnotesize{Hyper-parameters: KL divergence limit $\delta_{KL}$, backtracking coefficient $\alpha$, maximum number of backtracking
 steps $n_B$, behavioral memory size $M$, GAE lambda $\lambda\in(0,
 1]$, number of transitions $N$
 \State Input: initialize policy parameters
 $\boldsymbol{\theta}_0$, initial value function parameters
 $\boldsymbol{\omega}_0$
 }
 \For{$k=0, 1, 2, \ldots L$}
 \State Collect $N$ transitions $(\boldsymbol{s}_t, \boldsymbol{a}_t, r_t,
 \boldsymbol{s}_{t+1})$ by running policy ${\pi}$
  \State Set
  $\widehat{\boldsymbol{R}}=\boldsymbol{0}$ and $\widehat{\boldsymbol{A}}=\boldsymbol{0}$
  \For{$t=N-1,\ldots, 1, 0$}
     \begin{equation}\label{TD}
  \begin{cases}
    \widehat{\boldsymbol{R}}[t]=r_t + \gamma(1-d_t) \widehat{\boldsymbol{R}}[t + 1]       & \\
    \hat{\delta}=r_t+\gamma(1-d_t)
V_{\boldsymbol{\phi}}(\boldsymbol{s}_{t+1})-V_{\boldsymbol{\phi}}(\boldsymbol{s}_{t})
& \\
\widehat{\boldsymbol{A}}[t]=\hat{\delta} + \gamma\lambda(1-d_t)
\widehat{\boldsymbol{A}}[t+1] &
  \end{cases}
\end{equation}
  \EndFor
 \State Estimate the policy gradient
 \begin{equation}\label{gradient_app}
\hat{g} =
\frac{1}{N}\sum\limits_{t=0}^{N-1}\nabla_{\boldsymbol{\theta}_k}\log{\pi}_{k}
\widehat{\boldsymbol{A}}[t],
\end{equation}
\State Use the conjugate gradient algorithm to compute
$\hat{\boldsymbol{x}}_k=\hat{F}_k^{-1}\hat{\boldsymbol{g}}$
 \State Update the policy parameters:
 \begin{equation}\label{policy_param_update}
\boldsymbol{\theta}_{k+1} = \boldsymbol{\theta}_k +
\alpha^j\sqrt{\frac{2\delta_{KL}}{\hat{\boldsymbol{x}}_k^T\hat{F}_{\boldsymbol{\theta}_k}^{-1}\hat{\boldsymbol{x}}_k}}\hat{\boldsymbol{x}}_k,\,\,\,j=\{0,1,2,
\ldots, K\}
\end{equation}
\State Update the value network (via gradient descent)
\begin{equation}\label{value_net_update}
\boldsymbol{\omega}_{k+1} =
\mathrm{argmin}_{\boldsymbol{\omega}}\frac{1}{N}\sum_{t=0}^{N-1}\left(V_{\boldsymbol{\omega}}(\boldsymbol{s}_t)-\widehat{\boldsymbol{R}}[t]\right)^2.
\end{equation}
 \EndFor
\end{algorithmic}
\end{algorithm}

\emph{Updating Policy:} Updating policy is based on solving
optimization problem $\tilde{\mathcal{O}}$ which is done in
several steps (Step 5 to Step 10). First, we need to estimate the
rewards-to-go $\widehat{\boldsymbol{R}}$ and advantages
$\widehat{\boldsymbol{A}}$. In (\ref{TD}), $d_t\in \{0, 1\}$,
where $d_t=1$ implies that the episode is terminated. As a result,
the reward of the terminated time step of the episode is not
included in calculation of the advantages and rewards-to-go. On
the other hand, in the calculation of the advantages
$\widehat{\boldsymbol{A}}$ we adopt the generalized advantage
estimation (GAE) \cite{GeneralizedAdvantage} to improve the stability, where $\lambda\in(0,
1]$ is a given parameter.

\begin{figure*}[t]
\begin{center}$
\begin{array}{c}
\includegraphics[width=18cm, height=4cm]{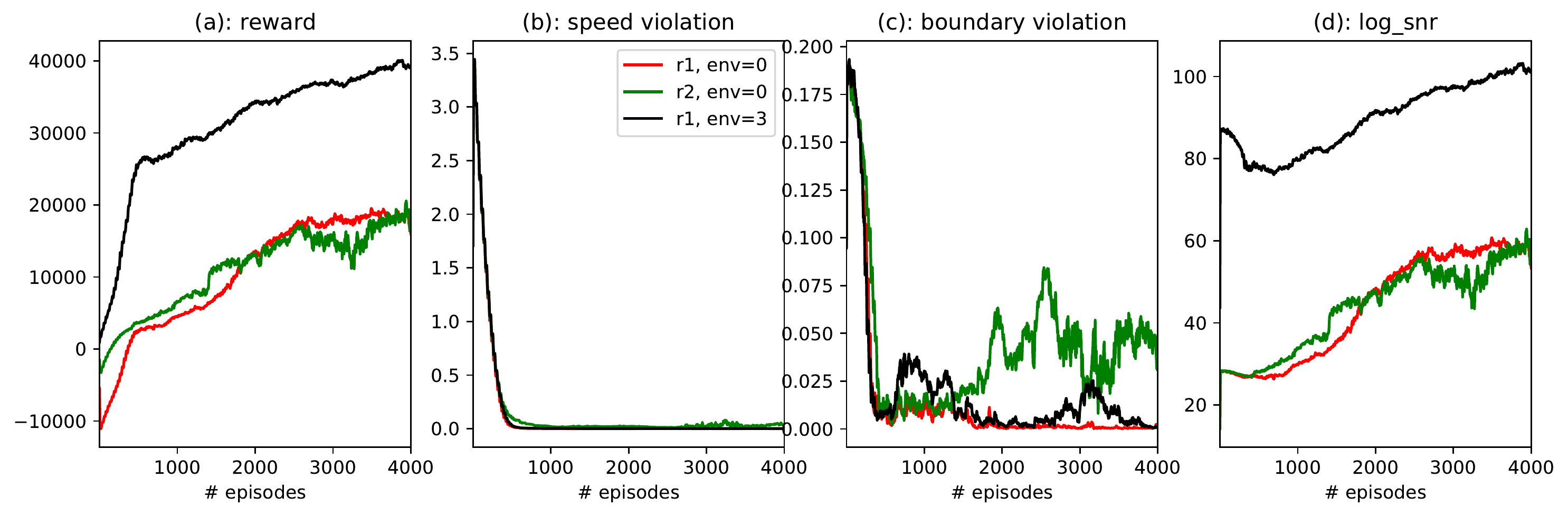}
\end{array}$
\end{center}
 \caption{(a): Average reward, (b): Average speed violation, (c): Average boundary violation of the search area, (d): Average logarithm of $\sum_k \frac{rss_k}{\sigma^2}$
(sum-RSS-to-noise ratio).  }
 \label{fig:reward}
\end{figure*}

The estimated advantages are then used to estimate the gradient
over the batch in Step 8. Steps 9 and 10 are to take the maximum
step for updating the current policy. First, in Step 9 we derive a
new direction via the conjugate gradient algorithm. Using
conjugate gradient algorithm one is able to solve
$\hat{F}_{\boldsymbol{\theta}_k}\hat{\boldsymbol{x}}_k=\hat{\boldsymbol{g}}$
through several iterations instead of resorting to the computation
of the inverse of FIM, hence substantially increasing the
computation efficiency and memory usage as the underlying DNN
could have millions of parameters. Step 10 known as line search is
a crucial step in TRPO algorithm as it ensures that the new
policy, which is derived based on the approximation of the
objective and the constraint, guarantees that actual surrogate
objective (not its linear approximation) is improved while the Kl
divergence constraint (not its quadratic approximation) stays
satisfied. In effect, the line search attempts to take possibly
the largest legitimate step toward the next policy. For a given
backtracking coefficient $\alpha<1$ the parameters
$\boldsymbol{\theta}_l$ are updated up to maximum backtracking
steps $J$. We terminate the line search when the smallest value
$\alpha^j$ (the bigger is $j$, the smaller will be the update
step) satisfies the KL divergence constraint and results in a
positive surrogate value.

\emph{Value Network:} The update of the value network
$V_{\boldsymbol{\omega}_k}$  is done in Step 11. Using the
rewards-to-go $\widehat{\boldsymbol{R}}$ the value network is
updated by mean-squared-error regression.

\begin{figure*}[t]
\begin{center}$
\begin{array}{c}
\includegraphics[width=18cm, height=4cm]{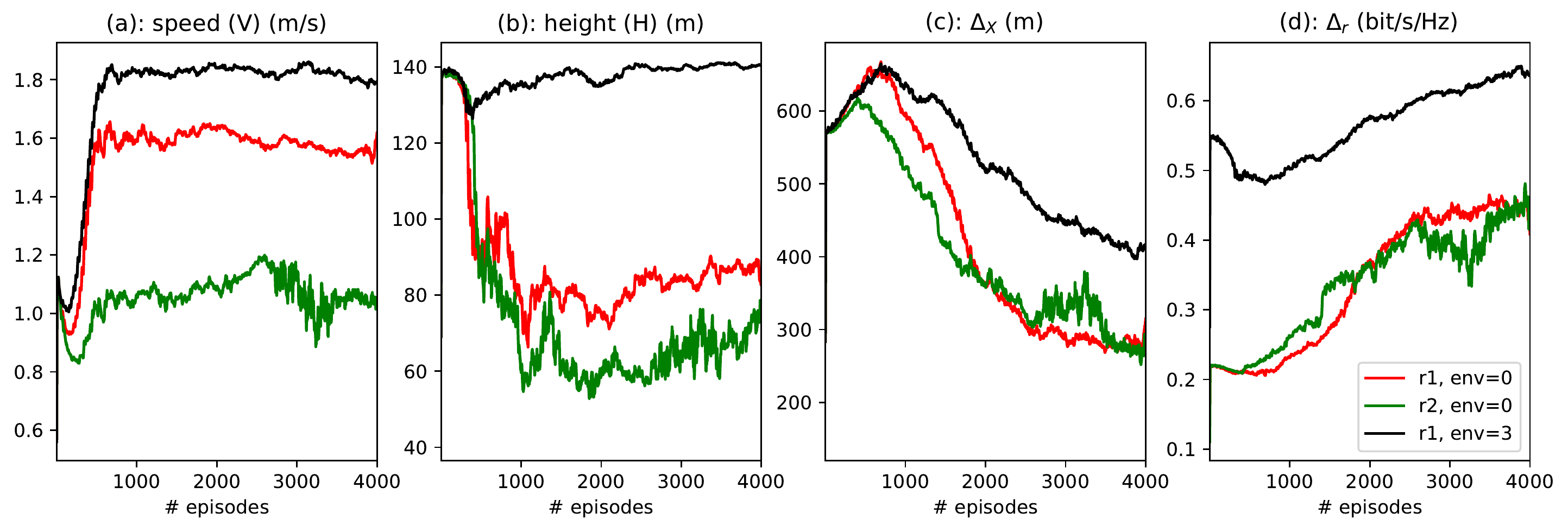}
\end{array}$
\end{center}
 \caption{(a): Average magnitude of speed $\boldsymbol{v}$, (b): Average height, (c): Average distance of UAV-BS to the center of the
 cluster, (d): Average $\Delta_r$. }
 \label{fig:rate}
\end{figure*}

\section{Experiments}
For the experiments we use the pytorch library
\cite{paszke2017automatic}. For each experiment we consider 6
different random seeds and calculate the average results
accordingly.

\subsection{Radio Environment}Now we discuss the communication
model of the environment that are used to produce RSS values and
transmitted data rates. We should emphasize that the provided information is
only used for numerical evaluations and are not known to the
agent. The UAV-BS is equipped with a directional antenna with beam-width of $w=\pi/3$. The
main-lobe and side-lobe antenna gains for UAV are $G=2.6/\omega^2$
and $g=G/100$. The locations of each user $k=1, 2,\ldots, K$ is
denoted by $(X_k, Y_k)\in \mathbb{R}^2\bigcap \mathcal{B}_C$,
where $\mathcal{B}_C$ stands for the cluster's geographical
boundaries. The vertical angel between the receiver $k$ and the
UAV-BS is $\rho_{k}=\tan^{-1}(H/\|X_{k}\|)$. The receiver $k$ is
within the main-lobe of the antenna, if $\rho_{k}>
\pi/2-\omega/2$, or equivalently for
$\|X_{k}\|<\frac{H}{\tan(\pi/2-\omega/2)}$ \cite{UAVantenna}. The
probability that the channel between UAV-BS and the receiver is in
LOS status is obtained from $p_{L}(\|X_{k}\|)=\Big(1+\phi e^{-\psi
\big(\frac{180}{\pi}\mathrm{arc}\tan(\frac{H}{\|X_{k}\|})-\phi\big)}\Big)^{-1}$
\cite{UAVchannel}, where $\phi$ and $\psi$ are the channel
parameters representing the characteristics of the communication
environment (see Table \ref{tableParams}). Note that the 3-D
distance between UAV-BS and user $k$ is $\sqrt{H^2+\|X_{k}\|^2}$.
The log-normal gain is also modelled via $\chi_{k}=10^{U_{k}/10}$
where $U_{k}\sim\mathcal{N}(\mu^l,\sigma^l_{k})$ in which
$\sigma^{l}_{k}=a_{l}e^{-c_{l}\frac{180}{\pi}\mathrm{arc}\tan(\frac{H}{\|X_{k}\|})}$
\cite{Al-Hourani} $a_{l}$ and $c_{l}$ are channel parameters (see
Table \ref{tableParams}). Furthermore, the fading power gain under
the LoS mode is modelled by Nakagami-m distribution with parameter
10. Under the NLoS mode the fading is modelled via unit-mean
exponential random variable. The background noise power is $-170$
dBm and the transmission power is 1 W. We also set the time slot
duration equals to 1 sec. We here assume that Doppler effect due
to the mobility of the UAV-BS as well as users is mitigated,
however, it is straightforward to include it in the simulations.

In the experiments, we consider two radio environments: $env=0$
(high-rise) and $env=3$ (sub-urban). We consider a circular search
area with radius 2000 meters and locate the cluster at position
$(1500, 1500)\in \mathbb{R}^2$. We assume the cluster is circular
with radius 100 meters. We then randomly locate 10 users in the
cluster. We also set $H_{\min}=40$ m, $H_{\max}=150$ m, $v_{\min}=0$ m/s, and $v_{\max}=100$ m/s.
 Note that users may dislocate in the cluster, but they
always stay in the cluster. We compared the delivered rate with a
\emph{heuristic approach} in which the agent knows the location of
the cluster and the locations of the users. The agent simply
locates itself in the middle of the cluster and chooses its height
such that all the users stay in the main-lobe of its antenna. We
then study the data rate ratio $\Delta_r$ that is the transmitted
data rate over the data rate achieved under the heuristic
approach.

\vspace{-.2cm}
\subsection{Policy and Value Networks} Policy is modelled
stochastically as a multivariate Normal distribution with diagonal
covariance matrix. The mean of this distribution is a DNN with 3
dens layers. The first and second layers are with input/output
dimensions $S/400$ and $400/300$ respectively, where $S$ is the
space dimension. This DNN has two heads, one for the mean value
and the other for the logarithm of the standard deviation. Each of
these are modelled by its associated dense layer with size $300/B$
where $B$ is the action dimension (number of users). Similarly,
the value net is also a DNN with three layers with the difference
that the last layer has dimensions $300/1$. The activation
functions are Tanh \cite{DeepLearning}. The state space is the
stacked received RSS values from all users. Regarding TRPO
algorithm, we set $\delta_{KL}=0.02$, $\lambda=0.94$, $N=10000$,
$L=4000$, $\gamma=0.99$, and $T=500$.

\begin{small}\begin{table}[t]
\centering \caption{Air-to-Ground parameters and the corresponding
values \cite{UAVchannel}.} \label{tableParams}
\begin{tabular}{|c|c c c c|}
 \hline
  & High-Rise & Dense-Urban & Urban & Sub-Urban \\ [0.5ex]
 \hline
 $\phi$ & 27.23 & 12.08 & 9.61 & 4.88 \\
 $\psi$ & 0.08 & 0.11 & 0.16 & 0.43\\
 $\mu_L$ & 1.5 & 1 & 0.6 & 0\\
 $\mu_N$ & 29 & 20 & 17 &18 \\
 $a_L$ & 7.37 & 8.96 & 10.39 & 11.25\\
 $a_N$ & 37.08 & 35.97 & 29.6 & 32.17\\
 $c_L$ & 0.03 & 0.04 & 0.05 & 0.06\\
 $c_N$ & 0.03 & 0.04 & 0.03 & 0.03\\[1ex]
 \hline
\end{tabular}
\end{table}\end{small}

\subsection{Impact of Reward Function}
Choosing a right form of the reward in navigation of UAV-BS is
complex given that the action, which carries out the navigation,
should be done based on RSS values while the actual goal is the
maximization of the transmission data rate. In this case, it is
not trivial to figure out how to optimally combine these components. Yet, we could compose the reward in the way that it
promotes the agent to take constructive actions while adheres to
physical limitations via imposing suitable penalties. For our experiments we consider two reward
functions: $r_2 = \sum_k (R_k + 0.01  \frac{rss_k}{\sigma^2}) - 5\Delta_a$ and
$r_1$:
\begin{equation}\label{r1}
r_1 =
  \begin{cases}
    0.1(\sum_k R_k + 0.01 \sum_k \frac{rss_k}{\sigma^2}) - 5\Delta_a      & \Delta_a>0\\
    \sum_k (R_k + 0.01 \frac{rss_k}{\sigma^2} ) & \Delta_a = 0\\
  \end{cases}
,
\end{equation}
where $R_k$ is the transmitted data
rate to user $k$. In both formulations the form of the reward
function promotes the movement toward receiving larger values for
RSS as well as delivering higher transmission data rate. As the
signal attenuations are highly affected by the path-loss
attenuation, which is a function of distance, we expect higher RSS
values correlate with higher transmission data rate (but this is
not guaranteed to take place due for instance to the effect of shadowing and
fading in the frequency multiplexed systems). Here, $\Delta_a$ is the sum of the penalties associated
with the feasible action (to enforce the constraints associated
with the magnitude of speed $[v_{\min}, v_{\max}]$, and azimuth
angel $[0, 2\pi]$, and polar angel $[0, \pi]$) and the search region
boundaries (to enforce the constraints regarding the altitude
$[H_{\min}, H_{\max}]$ and search area $(-2000\leq x \leq 2000,
-2000 \leq y \leq 2000$). As seen, compared to $r_2$, in $r_1$ the actual reward is scaled depending on whether the agent receives penalty
or not. This could discourage the agent
from unacceptable actions. However, via small rewards assigned to the violating actions
the agent is reminded that the actions were still constructive. In the formulation of $r_2$ such a
distinction is not provisioned, hence the agent may encounter difficulties to
distinguish beneficial actions out of heavy penalties.

We now
investigate which form of reward benefits the agent better in learning the
task. For this experiment, we consider high-rise environment ($env=0$).
From Fig. \ref{fig:reward}-a we observe that the agent gains
higher rewards under $r_2$ initially compared to $r_1$. However,
both of the rewards achieves almost the same average reward. On
the other hand, from Fig. \ref{fig:reward}-b we note that the
agent is able to learn the action boundaries very fast under both
reward forms. However, as we see from Fig. \ref{fig:reward}-c,
under the reward function $r_1$ the agent is able to more strictly
adhere to boundary limits of the search region compared to $r_2$.
Finally, Fig. \ref{fig:reward}-d shows that under both forms of
reward the agent is able to gather almost the same values of RSS.

From Fig. \ref{fig:rate}-a we see that the choice of reward
substantially affects the average speed of the agent. In effect,
under $r_1$ the agent tends to take higher speed values compared
to $r_2$. This might be due to the fact that the agent attempts to
correct its boundary violating actions (see Fig.
\ref{fig:reward}-c). This is also shown itself in the the height
of the agent under $r_1$. Fig. \ref{fig:rate}-c shows the distance
of the agent to the center of the cluster $\Delta_x$. We observe
that under both reward functions the agent learns to get closer to
the cluster center as a way to improve the transmission data rate.
We should note that the agent learns this behavior merely based on
RSS signals, which is interesting. Finally, in Fig.
\ref{fig:rate}-d we show the data rate ratio $\Delta_r$ under
both reward functions. We observe that under both rewards the
agent is able to improve its data rate. Interestingly, the agent is
able to achieve 40$\%$ of the heuristic scenario only based on RSS
values. We also note that both rewards are (almost) equally effective.

Consequently, while the choice of the reward does not have any
substantial impact on the transmitted data rate, distance to the
cluster, and action penalty, it has a profound impact on the speed
profile, the altitude of the agent, and how effectively the agent
is able to adhere to the search region boundaries. We therefore
consider $r_1$ in the rest of our experiments.

\subsection{Impact of Radio Environment}
Here, we attempt to demonstrate whether the agent is able to
recognize the impact of radio environment from the RSS values and
how she is responding to such a recognition. We consider two radio
environments $env=0$ (high-rise) and $env=3$ (sub-urban). Results
are shown in Fig. \ref{fig:reward} and Fig. \ref{fig:rate}. As
seen from Fig. \ref{fig:reward}-a, Fig. \ref{fig:reward}-d, and
\ref{fig:rate}-d, for an agent in $env=3$ the reward, average RSS
values, and the transmission data rate is much higher than
compared to the case of $env=0$. This is because in the former the
environment is more LOS dominant compared to the latter, hence the
signals go under less severe attenuations. The question is then
how the agent incorporates such recognition in its mobility?

From Fig. \ref{fig:rate}-a we observe that for the agent in
$env=3$ the magnitude of the speed is higher compared to the one
performing in $env=0$. Interestingly, the higher speed is used for
gaining much higher height (see \ref{fig:rate}-b). As a result,
the agent recognizes that for the radio environment with dominant
LOS component there is no need to get too close to the center of
the cluster if the height is properly adjusted. As seen, this
strategy can result in a decent rate transmission (about $60 \%$
of the rate in the heuristic scenario is achieved). On the other hand, for
$env=0$ the agent attempts to get closer to the cluster's center
(see \ref{fig:rate}-c) and simultaneously reduces its height
(\ref{fig:rate}-b) as an effective approach to circumvent
relatively higher path-loss.

\section{Conclusions}\label{Conclusions}
We addressed the mobility management of UAV-BS in a 3-D space to
support a cluster of users on the ground while the geographical
characteristics (e.g., location and boundary) of the cluster as
well as the geographical location of the users are not available. The
agent aimed at maximizing the data rate while the characteristics
of the radio environment are not known and may be extracted merely
from the received signal strength (RSS) from the users. We adopted
deep reinforcement learning to deal with the lack of model. In
particular, we adopted TRPO algorithm, which is an on-policy policy
gradient DRL, to adjust the (continuous) speed of UAV-BS only
based on RSS values. Our experiments suggested that the choice of
the reward substantially affects the speed profile and the ability
of the agent to adhere to its physical constraints. Interestingly,
we observed that UAV-BS was able to distinguish between high-rise
(less LoS dominant) and sub-urban (mainly LoS dominant)
environments.

\bibliographystyle{IEEEtran}
\bibliography{IEEEabrv,uav_rate}

\end{document}